\documentclass[12pt]{article}
\usepackage{epsfig}
\usepackage{psfrag}
\usepackage{amssymb}
\usepackage{latexsym}
\def\ps {SU(4)_C \times SU(2)_L \times SU(2)_R}

\def\be{\begin{equation}}
\def\ee{\end{equation}}
\def\bc{\begin{center}}
\def\ec{\end{center}}
\def\bea{\begin{eqnarray}}
\def\eea{\end{eqnarray}}

\def\nn{\nonumber}

\catcode`@=11
\def\marginnote#1{}
\newcount\hour
\newcount\minute
\newtoks\amorpm
\hour=\time\divide\hour by60
\minute=\time{\multiply\hour by60 \global\advance\minute by-\hour}
\edef\standardtime{{\ifnum\hour<12 \global\amorpm={am}%
        \else\global\amorpm={pm}\advance\hour by-12 \fi
        \ifnum\hour=0 \hour=12 \fi
        \number\hour:\ifnum\minute<10 0\fi\number\minute\the\amorpm}}
\edef\militarytime{\number\hour:\ifnum\minute<10 0\fi\number\minute}
\def\draftlabel#1{{\@bsphack\if@filesw {\let\thepage\relax
   \xdef\@gtempa{\write\@auxout{\string
      \newlabel{#1}{{\@currentlabel}{\thepage}}}}}\@gtempa
   \if@nobreak \ifvmode\nobreak\fi\fi\fi\@esphack}
        \gdef\@eqnlabel{#1}}
\def\@eqnlabel{}
\def\@vacuum{}
\def\draftmarginnote#1{\marginpar{\raggedright\scriptsize\tt#1}}
\def\draft{\oddsidemargin 0.0truein
        \def\@oddfoot{\sl preliminary draft \hfil
        \rm\thepage\hfil\sl\today\quad\militarytime}
        \let\@evenfoot\@oddfoot \overfullrule 3pt
        \let\label=\draftlabel
        \let\marginnote=\draftmarginnote
   \def\@eqnnum{(\theequation)\rlap{\kern\marginparsep\tt\@eqnlabel}%
\global\let\@eqnlabel\@vacuum}  }
\catcode`@=12
\begin{document}
\begin{titlepage}
\vspace*{-1cm}
\phantom{hep-ph/0501086} 
\hfill{DFPD-05/TH/18}

\vskip 0.5cm
\begin{center}
{\Large\bf Gauge coupling Unification and SO(10) in 5D}
\end{center}
\vskip 0.5  cm
\begin{center}
{\large Maria Laura Alciati}~\footnote{e-mail address: maria.laura.alciati@pd.infn.it}
~~~{\large and}
~~~{\large Yin Lin}~\footnote{e-mail address: yinlin@pd.infn.it}
\\
\vskip .1cm
Dipartimento di Fisica `G.~Galilei', Universit\`a di Padova 
\\ 
INFN, Sezione di Padova, Via Marzolo~8, I-35131 Padua, Italy
\\
\end{center}
\vskip 0.7cm
\begin{abstract}
\noindent
We analyze the gauge unification in minimal supersymmetric SO(10) grand unified theories in 5 dimensions.
The single extra spatial dimension is compactified on the orbifold $S^1/(Z_2\times Z_2')$ reducing
the gauge group to that of Pati-Salam $\ps$. The Standard Model gauge group is achieved by the further brane-localized
Higgs mechanism on one of the fixed points. There are two main different approaches developed in literature.
Higgs mechanism can take place on the Pati Salam brane, or on the $SO(10)$ preserving brane.
We show, both analytically and numerically, that in the first case a natural and succesfull gauge coupling unification can be achieved, 
while the second case is highly disfavoured. For completeness, we consider either the case in which the brane breaking
scale is near the cutoff scale or the case in which it is lower than the compactification scale.
\end{abstract}

\end{titlepage}
\setcounter{footnote}{0}
\vskip2truecm

%
%
\section{Introduction}
Grand unification is one of the most serious candidates for unification of particles and gauge interactions. 
Many properties of the Standard Model (SM) that seem mysterious or accidental, like the particle content, the cancellation of gauge
anomalies, the quantization of the electric charge, appear natural in the context of grand unified theories (GUTs). 
The minimal supersymmetric (SUSY) GUTs lead to a successful prediction of the weak mixing angle from unification of 
gauge coupling constants at the scale $M_U \sim 10^{16}$ GeV. Also the observed smallness of neutrino masses suggests 
the existence of right handed massive neutrinos with at least one having mass close to the GUT scale.
 
The conventional SUSY GUTs alone have many shortcomings that remain to be completed, such as the so-called doublet-triplet (D-T)
splitting problem and the too fast proton decay problem.  
An appealing mechanism to achieve the desired D-T splitting is when the grand unified symmetry is broken by the compactification 
mechanism in models with extra spatial dimensions \cite{kawa}. It has further been realized that SUSY GUTs formulated in five or 
more space-time dimensions give rise to many new prospectives. In such a framework, proton could be made stable by construction  
\cite{af,hebecker} or can proceed through dimension six operators \cite{hn1}. The $\mu$ problem is naturally absent when the gauge group
is broken by geometry of the extra-dimensions due to a continuous $U(1)_R$ symmetry \cite{MD,hn1,hn3}. 

The purpose of this work is to study the gauge coupling unification in five-dimensional SUSY SO(10) GUTs 
described in \cite{MD} and \cite{ KR}. In the context of GUTs based on the gauge group $SO(10)$, an entire generation
of quarks and leptons belongs to the same irreducible spinor representation together with a right handed neutrino.  
Predicting the existence of right handed neutrinos, $SO(10)$ provides a particularly interesting framework to realize 
the see-saw mechanism explaining naturally the smallness of neutrino masses. However, the breaking of SO(10) in five-dimensions 
(5D) to the SM group is more complicated than the breaking of SU(5) in 5D.
Since the reduction of grand unified symmetries by the type of orbifolding considered here does not reduce the rank
of the group \cite{hebecker2}, the breaking of SO(10) in 5D must be achieved by a combination of orbifold compactification and
conventional brane-localized Higgs mechanism. Two different approaches have been considerd in literature and their main 
difference deals with different localization of Higgs mechanism, that can takes place either on the PS brane \cite{KR} 
or on the SO(10) preserving brane \cite{MD}. 

We perform a detailed analysis in order to point out the impact of different choices of SO(10) breaking
on the unification of gauge coupling constants.
As guiding principles, we want to reproduce MSSM spectrum at low energies, while maintaining the nice feature of orbifold GUTs 
like the automatic D-T splitting. We find, both analitically and numerically, that the influence of 
the Higgs mechanism on the unification is not trivial and needs seriously to be taken under consideration.

In Sec. 2.1 and 2.2 we shortly review the orbifold construction and we focus on two different breaking chains in order to reduce SO(10) gauge group, namely Pati Salam breaking chain and SU(5) breaking chain. We notice that only the former can succesfully maintain the automatic D-T splitting and so we concentrate our studies on that. The further breaking of the Pati-Salam gauge symmetry to the SM gauge group can be accomplished via brane-localized Higgs machanism.
In Sec. 2.3 we carefully analyse the effects of the Higgs mechanism, considering the models \cite{MD} and \cite{ KR} already mentioned, denoting them as Pattern I and Pattern II respectively. For each of them we study all the possible values that can be assumed by the vacuum expectation value (VEV) of the Higgs field, focusing on two different regimes, that is high scale brane breaking (Case A) and brane breaking at an intermediate scale (Case B).
Then in Sec. 3, following \cite{AFLV}, we perform a detailed next-to-leading order analysis of gauge coupling unification, including two-loop running, heavy thresholds coming from KK particles, light thresholds from SUSY particles and SO(10) violating terms, due to the presence of kinetic terms at the PS brane, allowed in principle by the theory.
We pay a particular attention to the study of the heavy thresholds, highlighting the main difference between various patterns. It's well known that the renormalization group equations, including the two loop contributions, predict a value of the strong coupling constant $\alpha_3 (m_z)$ higher than nearby 9$\%$ of its experimental value. A more precise
gauge coupling unification can be obtained requiring an opposite contribution derived from KK particles 
in order to bring back $\alpha_3 (m_z)$ inside the experimental interval.
What we notice is that only for Pattern II and for the case of high scale brane breaking, the KK corrections have the correct sign.
In Sec. 4 we confirm our previus results by performing a more detailed numerical analysis and in Sec. 5 we conclude.
  
%

%
\section{SO(10) grand unification models in 5 dimensions}

\subsection{Orbifold construction}
We consider minimal supersymmetric SO(10) GUTs in 5 dimensions based on models constructed in \cite{MD, KR}. 
The 5-dimensional space-time is factorized into a product of the ordinary 4-dimensional space-time M$_4$ and 
of the orbifold $S^1/(Z_2\times Z_2')$, with coordinates $x^{\mu}$, ($\mu=0,1,2,3$) and $y=x^5$. 
The fifth dimension lives on a circle $S^1$ of radius $R$ with the identification provided by the two reflections: 
$ Z_2:y\rightarrow -y,$ and $Z_2':y'\rightarrow -y'$ with $y'\equiv y-\pi R/2$.
After the orbifolding, the fundamental region is the interval from $y=0$ to $y=\pi R/2$ with two inequivalent fixed points  
at the two sides of the interval. The origin $y=0$ and $y=\pi R$ represent the same physical point and similarly for
$y=+\pi R/2$ and $y=-\pi R/2$. When speaking of the brane at $y=0$, we actually mean the two four-dimensional slices at  
$y=0$ and $y=\pi R$, and similarly $y=\pi R/2$ stands for both $y=\pm\pi R/2$.

Generic bulk fields $\phi(x^{\mu},y)$ are classified by their orbifold parities $P$ and $P'$ defined by 
$\phi(x^{\mu},y) \to \phi(x^{\mu},-y)=P\phi(x^{\mu},y)$ and $\phi(x^{\mu},y') \to \phi(x^{\mu},-y')=P'\phi(x^{\mu},y')$.
We denote by $\phi_{\pm \pm}$ the fields with $(P,P')=(\pm,\pm)$ with the following $y$-Fourier expansions:
\bea
  \phi_{++} (x^\mu, y) &=& 
\sqrt{1 \over {2\pi R}}\phi^{(0)}_{++}(x^\mu)+
       \sqrt{1 \over {\pi R}} 
      \sum_{n=1}^{\infty} \phi^{(2n)}_{++}(x^\mu) \cos{2n y \over R}~~~,
\label{phi++exp}\nn\\
  \phi_{+-} (x^\mu, y) &=& 
       \sqrt{1 \over {\pi R}} 
      \sum_{n=0}^{\infty} \phi^{(2n+1)}_{+-}(x^\mu) \cos{(2n+1)y \over R}~~~,
\label{phi+-exp}\nn\\
  \phi_{-+} (x^\mu, y) &=& 
       \sqrt{1 \over {\pi R}}
      \sum_{n=0}^{\infty} \phi^{(2n+1)}_{-+}(x^\mu) \sin{(2n+1)y \over R}~~~,
\label{phi-+exp}\nn\\
 \phi_{--} (x^\mu, y) &=& 
       \sqrt{1 \over {\pi R}}
      \sum_{n=0}^{\infty} \phi^{(2n+2)}_{--}(x^\mu) \sin{(2n+2)y \over R}~~~.
\label{fourier}
\eea
where $n$ is a non negative integer. The Fourier component $\phi^{(n)}(x)$ of fields with opposite parities $(P,P')$
acquires a mass $(2n+1)/R$ upon compactification, while the component of fields with same parities acquires
a mass $(2n+2)/R$. As we will see, the structure of the even and odd Kaluza-Klein (KK) towers is crucial for the gauge unification. 
Only $\phi_{++}$ has a massless component and only $\phi_{++}$ and $\phi_{+-}$ are non-vanishing on the
$y=0$ brane. The fields $\phi_{++}$ and $\phi_{-+}$ are non-vanishing on the $y=\pi R/2$ brane, 
while $\phi_{--}$ vanishes on both branes.

\subsection{Gauge symmetry breaking}
\subsubsection{Pati Salam breaking chain}

The theory under investigation is invariant under N=1 SUSY in 5D, which corresponds to N=2 in four dimensions, 
and under SO(10) gauge symmetry. The gauge supermultiplet is in the adjoint representation of SO(10) and can be 
arranged in an N=1 vector supermultiplet $V$ and an N=1 chiral multiplet $\Phi$. We introduce a bulk Higgs hypermultiplet in
the fundamental representation of SO(10) which consists in two N=1 chiral multiplets $H_{10}$, $\hat H_{10}$ from 4-dimensional 
point of view. 

The parities of the fields are assigned in such a way that compactification reduces 
N=2 to N=1 SUSY and breaks SO(10) down to the PS gauge group $\ps$. The $P$ and $P'$ assignments are given 
in Table \ref{t1} \cite{MD, KR}.
The break down of N=2 to N=1 is quite simple 
and is achieved by the parity $P$. As illustrated in Table \ref{t1}, $H_{10}$ and $V$ have even $Z_2$ parities, while 
$\hat H_{10}$ and $\Phi$ have odd $Z_2$ parities and then vanish on the brane $y=0$. The additional parity $P'$ 
respects the surviving N=1 SUSY and can break the GUT gauge group. In fact, if we denote the PS and the SO(10)/PS 
gauge bosons as  $V^+$ and $V^-$ respectively, from the assignments of $Z_2'$ parities 
of Table \ref{t1} for $V^+$ and $V^-$, it turns out that, on the brane $y=\pi R/2$, only $V^+$ survives with the PS gauge
symmetry.

The projection $Z_2'$ can furthermore split the Higgs chiral multiplet $H_{10}$ ($\hat H_{10}$) in two chiral multiplets{\footnote{
The PS gauge group is isomorphic to the $SO(6) \times SO(4)$.}}: 
$H_{10}=(H_6, H_4)$($\hat H_{10}=(\hat H_{6},\hat{H}_{4}))$. $H_4$ contains scalar Higgs doublets $H^D_u$ and $H^D_d$ and $H_6$
contains the corresponding scalar triplets $H^T_u$ and $H^T_d$. As an important consequence of the parity assignments for
the Higgs fields in Table \ref{t1}, only the Higgs doublets and their superpartners are massless, while color triplets and
extra states acquire masses of order $1/R$, giving rise to an automatic D-T splitting.

Gauge symmetry would allow a mass term for the $H_{10}$ on the brane $y=0$ or a mass term for the $H_4$ ( and/or the $\hat H_{6}$ ) 
on the brane $y=\pi R/2$ as pointed out in \cite{MD}, thus spoiling the lightness of the Higgs doublets achieved by compactification,
but such a term can be forbidden by explicitly requiring an additional
U(1)$_R$ symmetry \cite{hn1,hn3}.
Therefore, before the breaking of the residual N=1 SUSY, the mass spectrum is the one shown in Table 1.

\begin{table}[h]
{\begin{center}
\begin{tabular}{|c|c|c|}   
\hline
& & \\                         
$(P,P')$ & field & mass\\ 
& & \\
\hline
& & \\
$(+,+)$ &  $V^+$, $H_4$ & $\frac{2n}{R}$\\
& & \\
\hline
& & \\
$(+,-)$ &  $V^-$, $H_6$ & $\frac{(2n+1)}{R}$ \\
& & \\ 
\hline
& & \\
$(-,+)$ &  $\Phi^-$, $\hat{H}_6$  
& $\frac{(2n+1)}{R}$\\
& & \\
\hline
& & \\
$(-,-)$ &  $\Phi^+$, $\hat{H}_4$ & $\frac{(2n+2)}{R}$ \\
& & \\
\hline
\end{tabular} 
\end{center}}
\caption{Parity assignment and masses ($n\ge 0$) of fields in the vector and Higgs supermultiplets. $V^+$ are PS gauge
bosons; $V^-$ are SO(10)/PS gauge bosons. $H_4$ contains the two MSSM scalar Higgs doublets while $H_6$ the corresponding 
Higgs triplets.}
\label{t1}
\end{table}

We notice that the broken gauge group at fixed points does not contain any $U(1)$ factors, so that the charge quantization is preserved.

\subsubsection{$SU(5)$ breaking chain}
The parities of the fields can be also assigned in such a way that compactification reduces N=2 to N=1 SUSY and breaks $SO(10)$ down to the gauge group $SU(5) \times U(1)$. Within this breaking chain, compactification would preserve SU(5) gauge symmetry, that is complete SU(5) multiplets survive after the orbifolding. So it's impossible to achieve an automatic doublet-triplet splitting, because doublets and triplets continue to belong to the same multiplet. We choose not to consider this case as we would lose one of the most attractive feature of models with extradimensions, namely avoiding to introduce an ad hoc and complicated scalar potential in order to explain the heaviness of Higgs triplets, while keeping the lightness of Higgs doublets.  
Another unpleasant feature of this breaking chain is the presence of the additional $U(1)$ factor that spoils the charge quantization. 

\subsection{Effects of brane Higgs mechanism}\label{branehiggs}

The breaking of Pati-Salam gauge symmetry to the SM gauge group can be accomplished via brane-localized 
Higgs mechanism. There are two different ways to realize Higgs mechanism on the brane in order to obtain 
the MSSM in the massless spectrum. The first way is to introduce a pair{\footnote {In
order to preserve supersymmetry two conjugate chiral Higgs fields are required and they need to acquire equal VEVs.}} 
of Higgs in the spinorial representation $(\bf{16})+(\bf{\bar{16}})$ of SO(10) living on the SO(10) preserving brane $y=0$. This breaks
the gauge group SO(10) down to SU(5) on the brane $y=0$. We will call this pattern of gauge symmetry breaking ``Pattern I''.
Alternatively the reduction of Pati-Salam gauge group into the SM gauge group can
be achieved directly on the symmetry breaking brane $y= \pi R/2$ by two Higgs in the $({\bf 4,1,2})+({\bf \bar4,1,2})$ of
$\ps$. This second possibility will be denoted as ``Pattern II''. 
To simplify, we will denote the Higgs field as $\Sigma$ either for the Pattern I or for the Pattern II. 
In all the two cases, we assume for $\Sigma$ a VEV along the right-handed neutrino direction $\nu_R$ and 
the resulting 4-dimensional theory has the SM gauge symmetry.

In order to preserve the gauge unification, one would expect that the SUSY Pati-Salam or SO(10) is broken on the brane
at the cutoff scale as traditional 4-dimensional SUSY GUTs \cite{intermediate}. In higher dimensions, KK states from bulk fields, charged under the SM 
gauge group, are important contributions for the gauge unification. For our purpose, 
to be as general as possible, we take the brane breaking scale $\langle \Sigma \rangle = u_{\Sigma}$ 
to be either near the cutoff scale $\Lambda$ where the gauge coupling is truly unified or at a scale less than 
the compactification scale $M_c$. We will explicitly show that the brane breaking cannot be achieved 
at intermediate scales $u_{\Sigma} \leq M_c$ including $u_{\Sigma}\sim M_c$.
\begin{figure}[h!]
\begin{center}
\includegraphics[width=9cm]{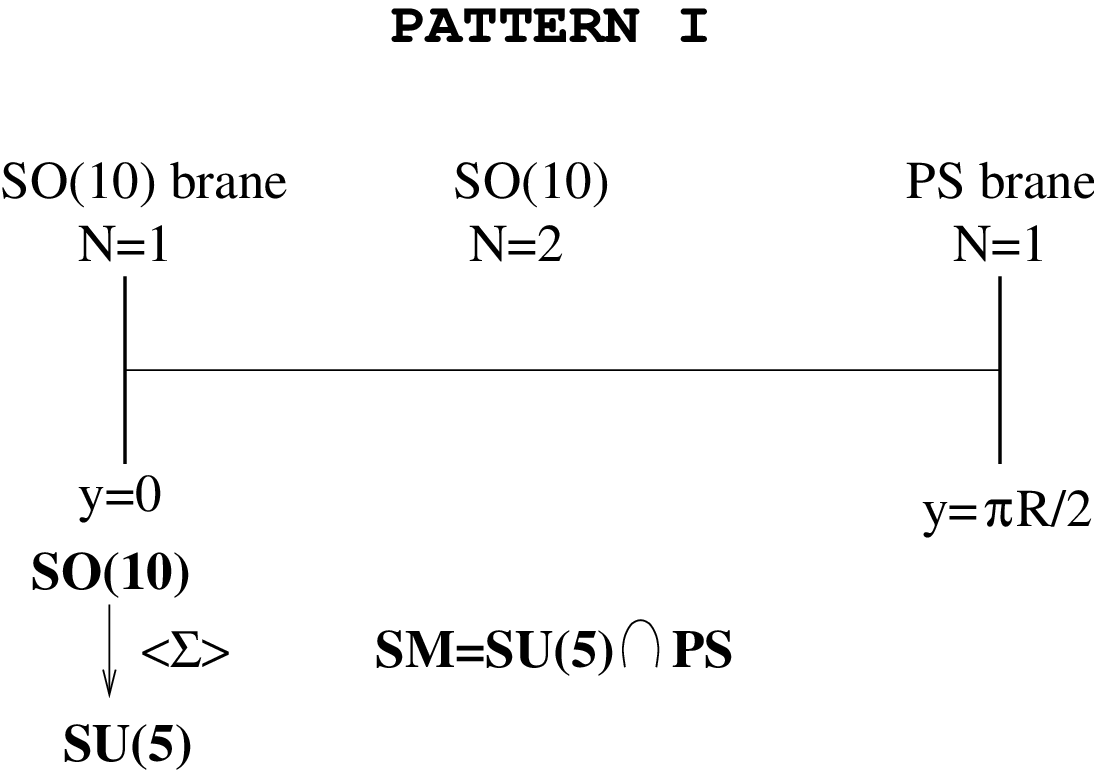}
\end{center}
\caption{A graphic reppresentation of Pattern I}
\label{pattern1}
\end{figure}

The breaking of the gauge symmetry by the brane field $\Sigma$ gives a
mass to the gauge fields localized on the brane \cite{Nomura:2001mf}, without affecting the spectrum of bulk Higgs fields. Precisely the Lagrangian has to be modified by the introduction of the following terms {\footnote{In order to preserve supersymmetry, similar 
terms are required also for gauginos.}}:

\be \rm{Pattern~I}: ~ {\cal L}~ \subset ~\delta(y) u_{\Sigma}^2  A_{\mu}^2, 
\qquad  \rm{Pattern~II}: ~{\cal L}~ \subset ~\delta\left(y-\frac{\pi R}{2}\right) u_{\Sigma}^2  A_{\mu}^2
\ee 
where $A_{\mu}$ are broken gauge bosons in SO(10)/SU(5) for Pattern I and that of PS/SM for Pattern II. 

\begin{itemize}
\item Pattern I

In this pattern, $A_{\mu}$ can be $(+,+)$ or $(+,-)$ with respect $Z_2 \times Z_2'$. To be explicit, $A^{(+,+)}_{\mu}$ 
are those bosons belonging to (SO(10)/SU(5))$\cap$PS~$=$~PS/SM and $A^{(+,-)}_{\mu}$ belong to (SO(10)/SU(5))$\cap$(SO(10)/PS)
~$=$~SO(10)/(SU(5)$\cup$PS){\footnote{These bosons are precisely the ones not present in SU(5), and they are $SU(2)_L$ doublets, color antitriplets with $Q=+2/3$ and $Q=-1/3$. They can also give rise to an effective $QQQL$ interaction, mediating proton decay.}}.
The wave functions are respectively 
\bea
A^{(+,+)}_{\mu}(x,y)&=&\sum_n N_n A^{(+,+)}_{\mu n}(x) \cos\left[M_n\left(\frac{\pi R}{2}-y\right)\right] \nn~;\\
A^{(+,-)}_{\mu}(x,y)&=&\sum_n N_n A^{(+,-)}_{\mu n}(x) \sin\left[M_n\left(\frac{\pi R}{2}-y\right)\right] \nn~,
\eea
where $N_n$ is normalization constant.
Accounting for the correct jumping conditions at $y=0$ the KK masses $M_n$ are given by:
\bea
\rm{for~} A^{(+,+)}_{\mu}: &&  M_n \tan \left(\frac{M_n \pi R}{2}\right) = \frac{g_5^2 u_{\Sigma}^2}2 \nn~; \\
\rm{for~} A^{(+,-)}_{\mu}: &&  M_n \cot \left(\frac{M_n \pi R}{2}\right) = \frac{g_5^2 u_{\Sigma}^2}2 \label{gbmassI}~.
\eea

\item Pattern II

\begin{figure}[h!]
\begin{center}
\includegraphics[width=9cm]{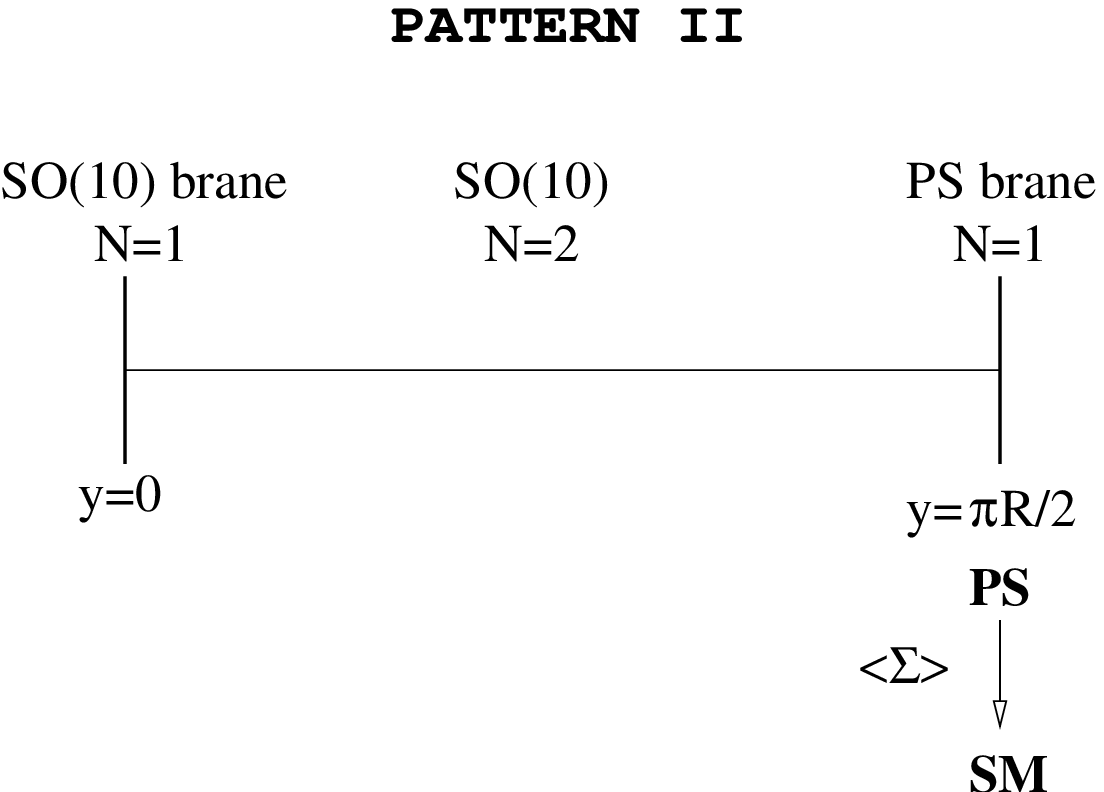}
\end{center}
\caption{A graphic reppresentation of Pattern II}
\label{pattern2}
\end{figure}

In the second pattern, differently, all gauge bosons $A _{\mu}$ have $(+,+)$ parity with respect to $Z_2 \times Z_2'$ and belong
 to PS/SM. The wave function is now more simple:
\bea
A_{\mu}=\sum_n N_n A_{\mu n}(x) \cos\left(M_n y\right) \nn~
\eea

and accounting for the correct jumping conditions at $y=\pi R/2$ the KK masses $M_n$ are given by:
\bea
 M_n \tan \left(\frac{M_n \pi R}{2}\right) = \frac{g_5^2 u_{\Sigma}^2}2 . \label{gbmassII}
\eea

\subsubsection{High scale brane breaking $\langle \Sigma \rangle \approx \Lambda$: Case A}
We first consider the case when the PS gauge group is broken on the brane near the cutoff scale of our theory, $\Lambda$.

\item Pattern I

When the PS gauge group is broken maximally near the cutoff scale $\Lambda$,
we have $g_5^2 u_{\Sigma}^2 R \gg 1$ and approximate solutions of the algebraic equations governing the KK masses
can be easily found in this limit. It's important to point out that the effect of the high scale localized mass term 
is to change the boundary condition for $A_{\mu}$ in such
a way that states with even KK mass are transfered to states with odd mass and vice versa. We find
\bea
\rm{for~} A^{(+,+)}_{\mu}: && M_n = \frac{2n}{R}  \rightarrow M_n  \simeq  \frac{2n+1}{R} \left(1-\frac1a\right), \\
\rm{for~} A^{(+,-)}_{\mu}: && M_n = \frac{2n+1}{R} \rightarrow M_n  \simeq  \frac{2n+2}{R} \left(1+\frac1a\right),
\eea
where $ a  =  \pi g_5^2 u_{\Sigma}^2 R/4 \gg 1$ and $n \geq 0 $. 

\item Pattern II

The jumping condition at $y=\pi R/2$ leads then to the following gauge boson spectrum in the limit 
$a  =  \pi g_5^2 u_{\Sigma}^2 R/4 \gg 1$:
\be M_n  \simeq  \frac{2n+1}{R} \left(1-\frac1a\right)~. \ee
A predictive framework of a gauge theory formulated in more than four dimensions 
requires generally that the theory is strongly coupled at the cutoff scale \cite{n2}.
Naive dimensional analysis with the strong coupling assumption gives $g_5^2 = 16\pi^3/\Lambda$ and $\langle\Sigma\rangle 
= \Lambda/4 \pi$. In order to account for the dependence of the gauge coupling unification on $u_{\Sigma}$, we introduce an 
new parameter $x$ defined by $u_{\Sigma}=x \Lambda/4 \pi$ with $x \leq 1$.
Then we can parametrize $a$ as  \be \label{defa} a =\frac {x^2 \pi^2}4 \frac {\Lambda} {M_c}~. \ee
\end{itemize}

\subsubsection{Brane breaking at a intermediate scale $\langle \Sigma \rangle \leq M_c$: Case B}
We now consider the regime $ u_{\Sigma} \ll M_c $. In this limit, 
either for the Pattern I or II, there are not any shifts between even and odd KK levels due to the brane breaking.
In the following, we will treat the Pattern I and II together.
The structure of even and odd KK masses remains approximately intact except for a small correction:
\bea
M_n = \frac{2n+1}{R} & \rightarrow & M_n  \simeq  \frac{2n+1}{R} \left[1+\frac b{(2n+1)^2\pi^2} \right],\label{gbmassvsodd} \\
M_n = \frac{2n+2}{R} & \rightarrow & M_n  \simeq  \frac{2n+2}{R} \left[1+\frac b{(2n+2)^2\pi^2} \right],\label{gbmassvseven}
\eea
where now  $b  =  \pi g_5^2 u_{\Sigma}^2 R/4 \ll 1$ and $n \geq 0 $.
Since the small correction $ \sim b/(n\pi)^2$ is smaller when $n$ increases, its contribution to the gauge coupling unification
is effectively neglected. What the Eqs.~(\ref{gbmassvsodd},~\ref{gbmassvseven}) do not yet account for
are the zero modes $A^{(+,+)}_{\mu0}$ ($\subset$PS) which, before the brane breaking, are all massless.
Unlike the case where the PS gauge group is broken maximally, here the subset of $A^{(+,+)}_{\mu0}$ belonging to PS/SM 
acquires a mass of order of $u_{\Sigma}$:
\be
M_0\approx \frac{g_5}{\sqrt{\pi R}}u_{\Sigma} = g_4 u_{\Sigma}\sim u_{\Sigma},
\ee
where $g_4$ is the effective 4-dimensional gauge constant.
In other words, when the PS gauge group is broken at an intermediate scale, the main effect of the localized Higgs mechanism
is to shift the zero-mode. Moreover, regarding the gauge coupling unification, as we will see, 
this shift of the zero-mode plays a very important role.

In the end, we are left with the regime $u_{\Sigma} \approx M_c $. In this case, the spectrum of the gauge bosons can only be obtained 
by resolving numerically the algebraic equations (\ref{gbmassI}) and (\ref{gbmassII}). Since we are interested in the gauge coupling unification
we will not report the solution here. Even though numerically more complicated, also for $u_{\Sigma} \approx M_c $,
we find that the leading contribution of the brane breaking to the spectrum of the gauge bosons is to give a mass of order of $M_c$
for $A^{(+,+)}_{\mu0}$ belonging to PS/SM. 

\subsection{Supersymmetry breaking}

The gauge coupling unification has a certain dependence on the light supersymmetric spectrum which arises from
the threshold corrections around the SUSY breaking scale. There are several SUSY breaking mechanisms that can be adapted 
to the present setup, but there is no compelling reason to prefer one to the other. It is possible to break SUSY by non-trivial
boundary conditions on the bulk superfields \cite{SSSB,SSSB1,bhn}.
SUSY can also be broken by an intrinsic four-dimensional
mechanism on either of the two branes. The soft SUSY breaking spectrum in MSSM must be very special in order to
avoid unacceptably large flavor violation. For this reason, the SUSY breaking sector and 
the transmission of SUSY breaking to the observable sector are notoriously a source of ambiguities
and phenomenological problems. For our purpose, since the gauge coupling unification is slightly affected by the light
SUSY particles, in order to parametrize our ignorance of the SUSY spectra, we will assume a variety of spectra as in \cite{AFLV}.
These spectra correspond to the so-called Snowmass Points and Slopes (SPS), a set of benchmark points and parameter 
lines in the MSSM parameter space corresponding to different scenarios \cite{SPS}. 
As we will see, this uncertainty is of a certain relevance but it is not the dominant one.


\section{Gauge coupling unification}

Following \cite{AFLV}, in this section we will provide a detailed analysis for the unification of gauge couplings.
The low-energy coupling constants $\alpha_i(m_Z)$ $(i=1,2,3)$
in the $\overline{MS}$ scheme are related to the unification scale 
$\Lambda_U$, the common value 
$\alpha_U=g_U^2/(4\pi)$ at $\Lambda_U$ and the compactification scale $M_c$
by the renormalization group equations (RGE):
\be
\frac{1}{\alpha_i(m_Z)}=\frac{1}{\alpha_U}
+\frac{b_i}{2\pi}\log \left(\frac{\Lambda_U}{m_Z}\right)
+ \delta^{NL}_i~.
\label{rge}
\ee
Here $b_i$ are the coefficient of the SUSY $\beta$ functions at one-loop,
$(b_1,b_2,b_3)=(33/5,1,-3)$, for 3 generations and 2 light Higgs SU(2) doublets. We recall that
$g_1$ is related to the hypercharge coupling constant $g_Y$ by
$g_1=\sqrt{5/3}~ g_Y$. Since the outcome of the gauge coupling unification does not depend on the universal contribution
of $\beta$ functions, we will employ the convention with $b_1=0$, so $b_i=(0,-28/5,-48/5)$.
In eq. (\ref{rge}), $\delta^{NL}_i$ stand for non-leading contributions and depend upon $M_c$. More precisely:
\be
\delta^{NL}_i=\delta^{(2)}_i+\delta^{(l)}_i+\delta^{(h)}_i+\delta^{(b)}_i~.
\label{delta}
\ee 
\begin{itemize}
\item $\delta^{(2)}_i$ represent two-loop running effects, coming from the gauge sector \cite{twoloop}:
\be
\delta_i^{(2)}=\frac{1}{\pi}\sum_{j=1}^3
\frac{b_{ij}}{b_j}
\log\left[1+b_j\left(\frac{3-8\sin^2\theta_W}{36\sin^2\theta_W-3}\right)\right]~.
\ee
\item
$\delta^{(l)}_i$ are light threshold corrections at the SUSY breaking scale\cite{susyth}:
\be
\delta_i^{(l)}=-\frac{1}{\pi}\sum_j b_i^{(l)}(j) 
\log\left(\frac{m_j}{m_Z}\right)~~~
\label{light}
\ee
where the index $j$ runs over the spectrum of SUSY particles of masses $m_j$
and extra Higgses. In the approximation where all particles have a common mass
$m_{SUSY}$ the SUSY contribution to coupling constants can be simply written as:
\be
\delta_1^{(l)}=-\frac{5}{4\pi}\log\frac{m_{SUSY}}{m_Z}~~~
\delta_2^{(l)}=-\frac{25}{12\pi}\log\frac{m_{SUSY}}{m_Z}~~~
\delta_3^{(l)}=-\frac{2}{\pi}\log\frac{m_{SUSY}}{m_Z}~~~.
\ee
The $\beta$-function coefficients $b_{ij}$ and $b^{(l)}_i$ can be found in \cite{AFLV}, see their Eq.~(4.6) and
Table 7 for some more details.
\item 
The contributions $\delta_i^{(b)}$ in (\ref{delta}) are originated by
kinetic terms for the gauge bosons of $\ps$ on the brane at $y=\pi R/2$. These terms, which break SO(10), are allowed
by the symmetries of the theory and, even if we set them to zero at the tree-level, they are generated by radiative corrections
\cite{cprt,ggh}. The brane kinetic terms arise from unknown ultraviolet physics above the cutoff scale $\Lambda$ and aim to 
modify the boundary value of the gauge coupling constants $g_i(\Lambda)$. 
Employing the strong coupling assumption, $g_i(\Lambda)$ receive mainly two contributions \cite{AFLV}:
\be
\frac{1}{g_i^2(\Lambda)}\approx\frac{\Lambda R}{8 \pi^2}+
O(\frac{1}{16\pi^2})~~~.
\label{glambda}
\ee
The first contribution comes from the SO(10)-invariant gauge coupling constant at the cutoff scale $\Lambda$ while the second one estimates the 
non universal contributions arising from the brane kinetic terms.
From (\ref{glambda}) we see that in order to rescue the predictivity of our 5-dimensional theory, such
as the gauge coupling unification, the SO(10)-symmetric component needs to dominate over the non-symmetric one.
Since it is welcome to keep the ultraviolet threshold contributions under control, in what follow we will assume $\Lambda R\gg1$.

Concretely, given our ignorance about the ultraviolet completion of our model, 
we will regard the threshold contributions $\tilde{\delta}_i^{(b)}$ 
from brane kinetic terms for the PS gauge bosons as random numbers with a flat distribution as in \cite{AFLV}. 
Observe that $\tilde{\delta}_i^{(b)}$ are different from $\delta_i^{(b)}$ because the latter are associated to the SM gauge group.
$\delta_i^{(b)}$ can be written easily in terms of $\tilde{\delta}_i^{(b)}$: 
\be
\label{deltaps}
\delta_1^{(b)}=\frac35\tilde{\delta}_1^{(b)}+\frac25\tilde{\delta}_3^{(b)}, \qquad
\delta_2^{(b)}=\tilde{\delta}_2^{(b)}, \qquad
\delta_3^{(b)}=\tilde{\delta}_3^{(b)}.
\ee

As a ``guide-line'' for our numerical computation, we will use the following ``natural'' interval for the contributions from
brane kinetic terms:
\be
\tilde{\delta}_i^{(b)}\in \left[ -\frac{1}{2\pi},+\frac{1}{2\pi}\right]~.
\label{random}
\ee
From (\ref{deltaps}), it's clear that $[-1/2\pi,1/2\pi]$ is the ``natural'' interval also for $\delta_i^{(b)}$.

\item
$\delta^{(h)}_i$ are heavy threshold corrections at the compactification scale $M_c$.
As becomes clear in the following, $\delta^{(h)}_i$  are the only contributions that feel
the effects of the different patterns of the gauge symmetry breaking.
\end{itemize}

\subsection{Next to leading effects}{\label{NLeffects}}

A successful model of GUTs should be well in agreement with the following experimental data \cite{pdg}:
\bea
\alpha_{em}^{-1}(m_Z)&=&127.906\pm 0.019\nn\\
\sin^2\theta_W(m_Z)&=&0.2312\pm 0.0002\nn\\
\alpha_3(m_Z)&=&0.1187\pm 0.0020~.
\label{inputrge}
\eea
It's well known that, at leading order, from the input values of $\alpha_{em}(m_Z)$ and $\sin^2\theta_W(m_Z)$ 
we obtain the prediction:
\be
\alpha_3^{LO}(m_Z)\approx 0.118~,
\label{lo}
\ee
compatible in an excellent level with the experimental value.
To investigate the effects up to the next to leading order, it's suitable to parametrize the  $\alpha_3(m_Z)$ in function of
the next to leading contributions $\delta^{NL}_i$ in the following way:
\bea
\label{deltas}
\alpha_3(m_Z)&=&\alpha_3^{LO}(m_Z)\left[1-\alpha_3^{LO}(m_Z)\delta_s\right]\nn\\
\delta_s&=&\frac{1}{7}\left(5\delta^{NL}_1-12\delta^{NL}_2+7\delta^{NL}_3\right)~.
\eea

Before analyzing numerically the evolution of RGE for the gauge coupling constants including all non-leading effects, 
we could investigate a qualitative prediction of the strong coupling constant $\alpha_3(m_Z)$, forgetting for a moment 
the heavy thresholds  $\delta^{(h)}_i$. From our previous discussion, we can already evaluate $\delta_s$ 
from the contributions of two-loop corrections and light thresholds independent from $M_c$:
\bea
\delta_s^{(2)}&\approx& -0.82\nn\\
\delta_s^{(l)}&\approx& -0.50+\frac{19}{28\pi}\log\frac{m_{SUSY}}{m_Z}~.\nn
\eea
Considering the SUSY spectra adopted for our numerical analysis, we find that 
the combined effect $\delta _s^{(2)}+\delta _s^{(l)}$ gives $\delta _s \lesssim -0.62$ (but typically 
$\delta _s \approx -0.79$) and would raise the prediction of $\alpha_3(m_Z)$ at least to approximately 0.127 
(but typically to 0.129). 

Since $(-1/2\pi,+1/2\pi)$ is ``the natural interval'' of the random contribution $\delta^{(b)}_i$, from (\ref{deltas}), we have $\delta_s^{(b)} \leq \frac1{2\pi}\frac17(5+12+7) \approx 0.55$. 
Including all contributions independent on the compactification scale together, $\delta_s$
is then always negative. Consequently, in order to bring back $\alpha_3(m_Z)$ inside the experimental interval, 
or the correction $\delta^{(h)}_s$ from heavy thresholds must be positive or we have to unnaturally enlarge the
random interval of $\delta^{(b)}_i$. As we will see later, 
the sign of $\delta^{(h)}_s$ can be positive or negative depending on whether the PS gauge group is
broken on the brane by Pattern I or II. The sign of $\delta^{(h)}_s$ depends also on the scale of
the localized Higgs VEV $\langle \Sigma \rangle$, that is Case A or B.
From this point of view, the request for the precise and natural unification of the 
gauge couplings allows us to select the pattern of the gauge symmetry breaking of SO(10) in 5D.

\subsection{Heavy Thresholds}{\label{heavy}}

The heavy thresholds are strongly model-dependent and affect the desired
gauge coupling unification. In this section we will give an extensive study of the heavy threshold effects
for the various patterns of the gauge symmetry breaking mechanism on brane discussed in Sec.~(\ref{branehiggs}).
The aim is to calculate the next-to-leading effect on the prediction of the strong coupling constant $\alpha_3(m_Z)$ 
given by (\ref{deltas}) due to the heavy threshold contributions. As already announced from our qualitative analysis
in Sec. (\ref{NLeffects}), a successful gauge coupling unification favors patterns with positive $\delta^{(h)}_s$.
To evaluate the heavy threshold effects, we use the leading logarithmic approximation for the particles whose masses
are smaller than the cut-off scale $\Lambda$, that is
\be
\delta^{(h)}_i =
\frac{b_i}{2\pi}
\sum_{n} \log\frac{\Lambda}{M_n} ~.
\label{lla}
\ee
In (\ref{lla}), the sum is performed on all states belonging to the Kaluza-Klein towers of the gauge bosons 
and the Higgs fields. 

\subsubsection{Case A} \label{casea}

Starting with the case where the PS gauge group is broken on brane by a VEV of the cutoff scale $\Lambda$, we find from (\ref{lla})
that, for both pattern I and II, the heavy threshold contributions to RGE are given by the formula:
\be
\delta^{(h)}_i \approx
\frac{\alpha_i}{2\pi}
\sum_{n=0}^N \log\frac{(2n+2)}{(2n+1)}+
\frac{\beta_i}{2\pi}\frac{(2N+2)}{a}~,
\label{hta}
\ee
with the coefficients $\alpha_i$ and $\beta_i$ depending on the patterns and listed in Table~2. 
The sums stop at $N$ so that $(2N +2) M_c$ is the Kaluza-Klein level closest to, but still
smaller than, the cutoff $\Lambda$:
\be
(2 N+2)\approx\frac{\Lambda}{M_c}~.
\ee
For large $N$, that is for $\Lambda R\gg 1$
\be
\label{approx}
\sum_{n=0}^N \log\frac{(2n+2)}{(2n+1)}
\approx
\frac{1}{2}\log(N+1)+\frac{1}{2}\log\pi
\approx
+\frac{1}{2}\log\frac{\Lambda}{M_c}+\frac{1}{2}\log\frac{\pi}{2}~.
\ee
In this limit where the strong coupling condition is verified, we can use the expression (\ref{defa}) and 
the heavy thresholds (\ref{hta}) become:
\be
\delta^{(h)}_i \approx
\frac{\alpha_i}{4\pi}\log\frac{\Lambda}{M_c}+\frac{\beta_i}{2\pi}\frac4{x^2 \pi^2}+...
~~~,
\label{htaapprox}
\ee
where dots stand for universal contributions. We insist on the fact that, up to an irrelevant universal contribution redefining
the initial condition $\alpha_U$, all the effect comes from the shift between even and odd Kaluza-Klein levels 
that removes the degeneracy within full SU(5) and SO(10) multiplets.

\begin{center}
\begin{tabular}{|c||c|c||c||c|c|}   
\hline
& & & & &\\
$\alpha_i$& $\rm{I}$ & $\rm{II}$ & $\beta_i$& $\rm{I}$ & $\rm{II}$ \\
& & & & &\\
\hline
\hline
& & & & &\\
1& 0& 0 & 1&0 & 0\\
& & & & &\\
\hline
& & & & &\\
2&$\frac{44}5$ & $\frac{16}5$ & 2&$\frac{28}5$ & $\frac{14}5$\\
& & & & &\\
\hline
& & & & &\\
3&$\frac{54}5$ & $\frac{36}5$ & 3& $\frac{18}5$ & $\frac{9}5$\\
& & & & &\\
\hline
\end{tabular}
\end{center}
\vspace{3mm}
Table 2: $\alpha_i$ and $\beta_i$ defined in (\ref{hta}) for the patterns I and II. 
The gauge symmetry breaking on brane is achieved by a cutoff scale VEV.
\vskip 0.5cm

From (\ref{htaapprox}) and using the coefficients given in Tab.~2, it's easy to understand qualitatively
the KK contributions for $\alpha_3(m_Z)$:
\bea
\delta^{(I)}_s &=& \frac{-15}{14\pi}\log\frac{\Lambda}{M_c}-\frac{12}{\pi^3}\frac1{x^2}~,\label{deltasIa}\\
\delta^{(II)}_s &=& \frac{3}{7\pi}\log\frac{\Lambda}{M_c}-\frac{6}{\pi^3}\frac1{x^2}~\label{deltasIb}.
\eea
For the Pattern I, $\delta^{(I)}_s$ is always negative and consequently the heavy thresholds would further raise 
the prediction of $\alpha_3(m_Z)$ as explained previously. In this case 
the compactification scale is necessarily very close to the cutoff scale, both similar to the unification scale in
conventional GUTs.
However, as already discussed before, $\Lambda R \gg 1$ is assumed in order to garantee the dominance of SO(10)-symmetric
bulk contributions over the non symmetric brane contributions.
Furthermore Eq.~(\ref{glambda}) predicts, in the strong coupled case, a hierarchy of order
$(4\pi)^2$ between the cutoff scale and the compactification scale.
As a conclusion, the Pattern I is highly disfavored regarding the gauge coupling unification. 
Contrarily, in the Pattern II, the contribution proportional to $\log\Lambda/ M_c$ has the right sign 
with a relatively small coefficient. As a result, if we neglect the dependence on $x$, a considerable large gap $\Lambda /M_c \gtrsim 100$ is required
to reproduce the experimental value of $\alpha_3(m_Z)$. This is very welcome from the view-point of the the predictivity
of our theory.

\subsubsection{Case B} {\label{caseb}}

We follow then with the case of the brane breaking by an intermediate scale VEV, the leading heavy threshold contributions 
to RGE can be found analytically only in the limit $ u_{\Sigma} \ll M_c $:
\be
\delta^{(h)}_i \approx
\frac{\alpha_i}{2\pi}
\sum_{n=0}^N \log\frac{(2n+2)}{(2n+1)}+
\frac{\beta_i}{2\pi}\log\frac{\Lambda}{u_{\Sigma}}~,
\label{htb}
\ee
where we have neglected the small corrections $ \sim b/(n\pi)^2$ to the spectrum of gauge bosons given in
Eqs.~(\ref{gbmassvsodd},~\ref{gbmassvseven}). In this case, the result formula (\ref{htb}) is the same for
both patterns I and II with the same coefficients $\alpha_i$ and $\beta_i$.

For large $N$, we can use the approximation (\ref{approx}) and then, put $u_{\Sigma}=y M_c$ with $y\ll 1$,
the expression (\ref{htb}) becomes
\be
\label{htbapprox}
\delta^{(h)}_i \approx 
\frac{\sigma_i}{4\pi}
\log\frac{\Lambda}{M_c}+
\frac{\beta_i}{2\pi}\log \frac{1}{y}+...
\ee
Curiously, the coefficients $\sigma_i$ are exactly the same of $\alpha_i$ in Tab.~2 for the Pattern I.

\begin{center}
\begin{tabular}{|c||c|c||c||c|c||c|c|c|}   
\hline
& & & & & & & &\\
$\alpha_i$& $\rm{I}$ & $\rm{II}$ & $\beta_i$& $\rm{I}$ & $\rm{II}$ & $\sigma_i$& $\rm{I}$ & $\rm{II}$ \\
& & & & & & & &\\
\hline
\hline
& & & & & & & &\\
1& 0& 0 & 1&0 & 0&1 &0 &0\\
& & & & & & & &\\
\hline
& & & & & & & &\\
2&$-\frac{12}5$ & $-\frac{12}5$ & 2&$\frac{28}5$ & $\frac{28}5$ & 2 & $\frac{44}5$ & $\frac{44}5$ \\
& & & & & & & &\\
\hline
& & & & & & & &\\
3&$\frac{18}5$ & $\frac{18}5$ & 3& $\frac{18}5$ & $\frac{18}5$ & 3 & $\frac{54}5$ & $\frac{54}5$ \\
& & & & & & & &\\
\hline
\end{tabular}
\end{center}
\vspace{3mm}
Table 3: $\alpha_i$, $\beta_i$ and $\sigma_i$ defined in (\ref{htb}), (\ref{htbapprox}) for the patterns I and II. 
The gauge symmetry breaking on brane is achieved by a scale $\langle \Sigma \rangle\ \ll M_c$.
\vskip 0.5cm

We can then proceed with the calculation of the $\delta^{(h)}_s$ in order to point out the heavy threshold contributions for 
$\alpha_3(m_Z)$. The result is the same for both patterns I, II:
\be
\delta^{(h)}_s=\delta^{(I)}_s=\delta^{(II)}_s = \frac{-15}{14\pi}\log\frac{\Lambda}{M_c}-\frac{3}{\pi}\log \frac1{y}~.
\ee
The logarithmic contribution $-15/14\pi\log(\Lambda/M_c)$ is identical to that obtained in Eq.~(\ref{deltasIa})
and the term proportional to $\log (1/y)$ can give a further large negative contribution to $\delta^{(h)}_s$ since
$y \ll 1$. Then the limit $\langle \Sigma \rangle\ \ll M_c$ is really ruled out regarding the gauge coupling unification.

The previous conclusion is not essentially changed when we move to the regime $\langle \Sigma \rangle\ \approx M_c$.
For this regime, in fact, we can employ the numerical solutions of the algebraic equations (\ref{gbmassI}) and (\ref{gbmassII}) for the gauge boson spectra.
We obtain subsequently the numerical dependence of $\delta^{(h)}_s$ on $N \sim \Lambda/ M_c$ and find that $\delta^{(h)}_s$
is negative anyway for both patterns I, II. Then the Case B, where the brane-localized Higgs gives a VEV less 
than the compactification scale, is also strongly disfavored from the viewpoint of the unification of the gauge coupling constants. 


\section{Numerical results}
In this section we will perform a more detailed numerical study in order to confirm the preliminary analysis given
in sections (\ref{NLeffects}) and (\ref{heavy}). We use the experimental inputs shown in eq.~(\ref{inputrge}) 
and the non-leading contributions $\delta^{NL}_i$ given in eq.~(\ref{delta}) to solve the renormalization group
equations (\ref{rge}). It's clear that physical solutions shall
satisfy the condition $\Lambda/M_c>1$, being $\Lambda$ the cutoff scale. We remember that only for the Pattern II of the Case A 
we have obtained a positive $\delta^{(h)}_s$ and in all the other cases $\delta^{(h)}_s < 0$. 
We will analyze the Pattern I and II separately.

\begin{itemize}
\item Pattern I

From the previous analytical analysis and the numerical inspection, we find that, either for Case A or Case B,
$\delta^{(h)}_s < -15/14\pi\log(\Lambda/M_c)$. As a guide line, it's sufficient to consider only the heavy thresholds 
proportional to $\log(\Lambda/M_c)$. Even in this most favored situation for the gauge coupling unification,
we can not find any numerical solutions at all with $\Lambda/M_c>1$ for Pattern I if we consider 
the ``natural'' interval: $b_i^{(b)} \in (-1/2\pi,+1/2\pi)$. 

In principle, the gauge coupling unification can be restored if we increase the contribution from the brane kinetic terms 
to intervals larger than $(-1/2\pi,+1/2\pi)$. But since the mean contribution of the random values of $b_i^{(b)}$ is zero, 
this indicates that, for the pattern under consideration, the solutions of RGE, if they exist, must be highly fine-tuned.
In fact, we see from the scatter plots Fig.~\ref{mafi} that, even using a larger interval $(-2/\pi,+2/\pi)$, 
the solutions are obtained for $\Lambda/M_c>1$ only in $7\%$ of the random numbers of $\delta^{(b)}_i$, 
that is with very special combinations of $\delta^{(b)}_i$.
\begin{figure}[h!]
\begin{center}
\includegraphics[width=7.9cm]{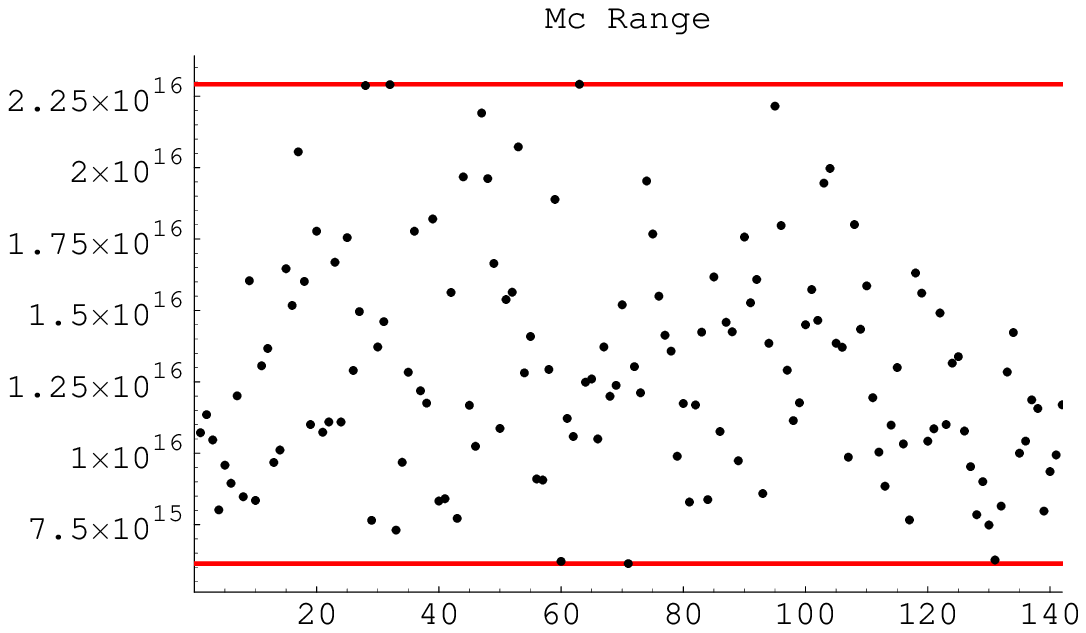}
\includegraphics[width=7.9cm]{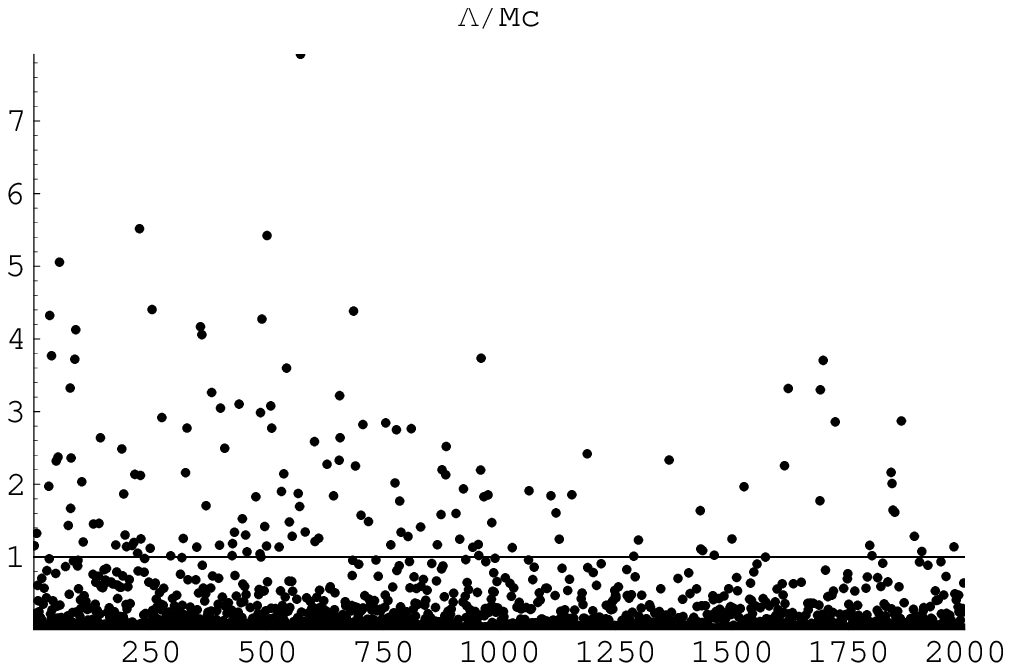}
\end{center}
\caption{The random samplings of the compactification scale $M_c$ and the ration $\Lambda/M_c$ for 
 Pattern I. Level of fine-tuning: 7 $\%$. The physical solutions satisfy $\Lambda/M_c > 1$.The range of the contributions from the SO(10)-breaking brane terms 
is $\delta_i^{(b)}\in \left[-2/\pi,+2/\pi\right]$. The horizontal lines in the first figure correspond to the maximum and the minimum values of $M_c$.}
\label{mafi}
\end{figure}

Since we want to construct a predictive theory, it's hopeful that the SO(10) symmetric contribution to 
the gauge coupling constants dominates over the SO(10) non-symmetric one. 
For this reason, the growth of $\delta^{(b)}_i$ leads a corresponding growth of $\Lambda R$ in order to keep 
under control the brane kinetic contribution. Nevertheless, as shown in the scatter plots Fig.~\ref{mafi}, 
the numerical solutions have likely low values of $\Lambda / M_c$ ( $\Lambda / M_c <10$). 
Consequently the highly fine-tuned solutions are not acceptable anyway.

\item Pattern II

In the case with high VEV scales, Case A, there are numerical solutions even if we turn off the brane kinetic contributions 
$\delta_i^{(b)}$. The brane kinetic terms introduce a theoretical uncertainty on the determination of the compactification scale.
We have used Eq.(\ref{htaapprox}) to perform our numerical solutions of RGE and the
Fig.\ref{rabi1} and Fig.\ref{rabi0.5} show the compactification scale $M_c$ and the ration $\Lambda/M_c$ versus SUSY spectrum for the Case A 
with 2 different values of $x$. The dominant errors come from the unknown SO(10) violating brane interactions parametrized by
the random distribution $\delta_i^{(b)}\in (-1/2\pi,+1/2\pi)$. 
From the graphics of the ration $\Lambda/M_c$, we can see that the average value of $\Lambda/M_c$ for $x=1$
is about 1000 and it depends very sensitively on the parameter $x$. 

In the other regime, $\langle \Sigma \rangle \leq M_c$, that is Case B, the situation changes completely because 
$\delta^{(h)}_s < 0$. We have shown numerically that this case is essentially similar to that of Pattern I
and we will not reproduce the scatter plots for this case.
For the most extreme regime where $\langle \Sigma \rangle \ll M_c$, we have already shown analytically in Sec.~(\ref{caseb})
that Pattern II is ruled out. Moving to $\langle \Sigma \rangle \approx M_c$, the numerical check shows that
$\delta^{(h)}_s > -15/14\pi\log(\Lambda/M_c)$ for low values of $\Lambda/M_c \lesssim 100$ but the sign is always negative, then the conclusion is unchanged. Then Case B is strongly disfavored 
for Pattern II.

\begin{figure}[h!]
\begin{center}
\includegraphics[width=7.9cm]{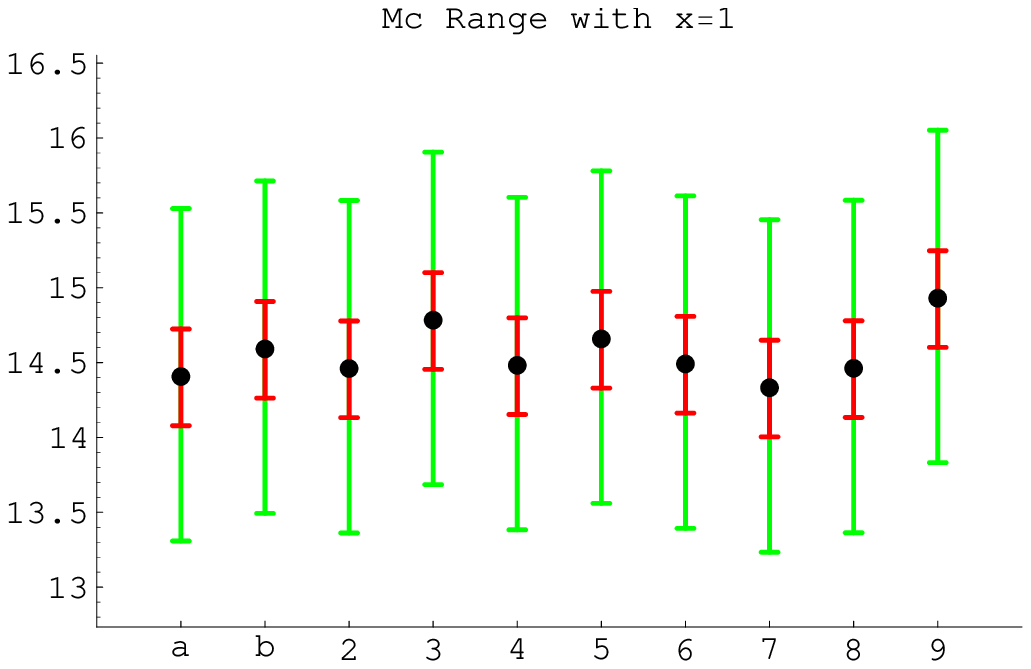}
\includegraphics[width=7.9cm]{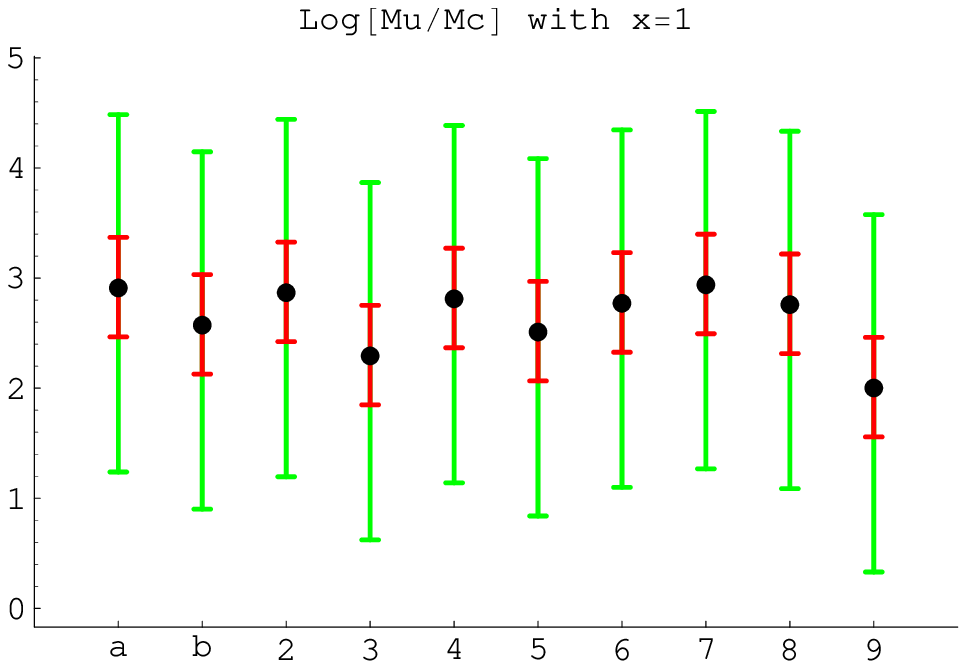}
\end{center}
\caption{Compactification scale $M_c$ and the ration $\Lambda/M_c$ versus SUSY spectrum for the Case A of the Pattern II 
with $x=1$. The shorter error bar represent the parametric error dominated by the experimental
uncertainty on $\alpha_3(m_Z)$, the wider bar includes the dominant source 
of error, the SO(10)-breaking brane terms $\delta_i^{(b)}\in \left[-1/2\pi,+1/2\pi\right]$.}
\label{rabi1}
\end{figure}
\begin{figure}[h!]
\begin{center}
\includegraphics[width=7.9cm]{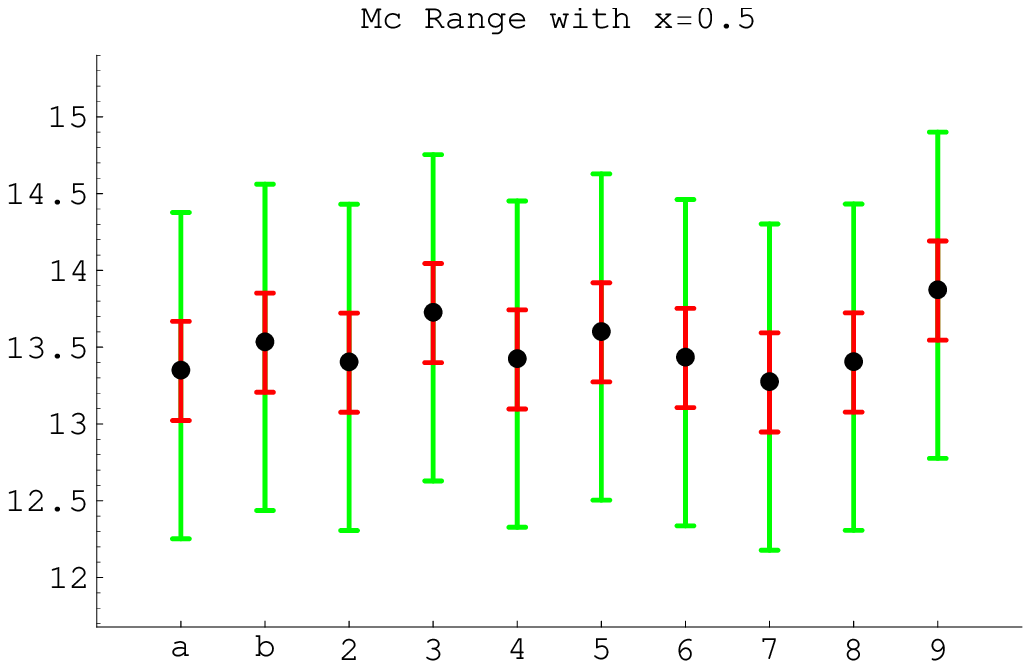}
\includegraphics[width=7.9cm]{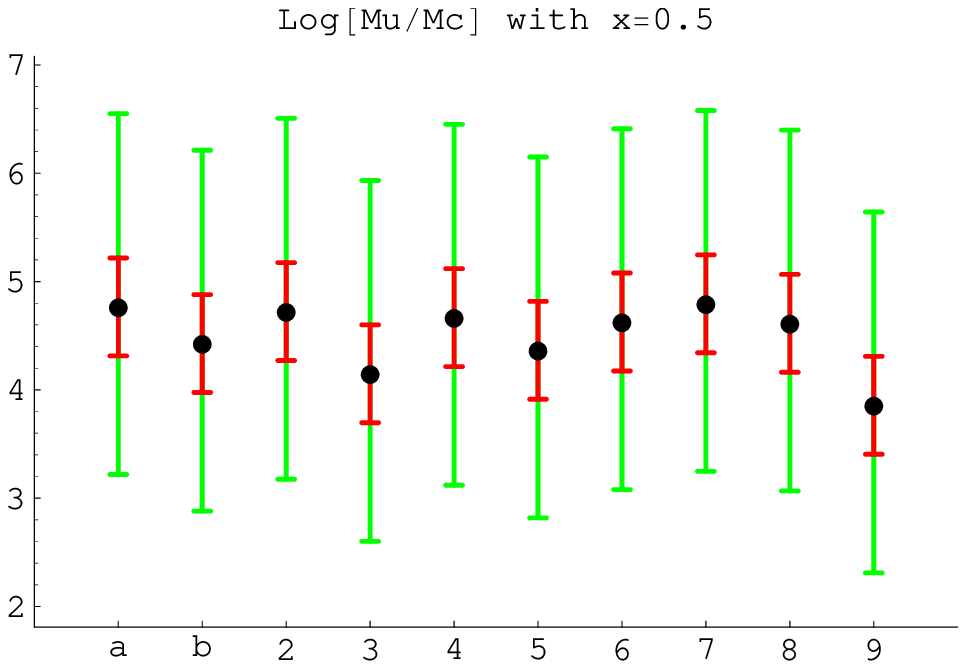}
\end{center}
\caption{Compactification scale $M_c$ and the ration $\Lambda/M_c$ versus SUSY spectrum for the Case A of the Pattern II 
with $x=1/2$. Error bars as in Fig.~{\ref{rabi1}}.}
\label{rabi0.5}
\end{figure}
 
\end{itemize}


\section{Conclusion}

SUSY GUTs with extradimensions based on SO(10) gauge group, not only mantain the beautiful feature of unifying all matter including right handed neutrino in the same gauge multiplet, but also overcome some problems derived from the difficulty to explain the lightness of Higgs doublets and the heavyness of Higgs triplets, both occuring in the same gauge multiplet, without the introduction of a complicated 
or fine-tuned scalar sector, as happened in four-dimensional GUTs.
The complete breaking of SO(10) gauge group, with rank equal to 5, to the SM gauge group, with rank 4, requires both orbifold and Higgs mechanisms. The latter can proceed in two different ways, but preferring one to another is not a trivial choice, as it has a great impact on the gauge coupling unification as we have showed in this paper.

Conventional four-dimensional SUSY GUTs predict in general a value of the strong coupling constant $\alpha_3 (m_z)$ 
higher than the largest value allowed by its experimental error. In 5D, it's possible to improve the precision of the 
gauge coupling unification including the contribution derived from KK particles since this heavy threshold contribution 
could bring back $\alpha_3 (m_z)$ inside the experimental interval. The correct sign of 
heavy thresholds depends on the patterns used to break the SO(10) gauge symmetry to that of the SM.
If we want to preserve the requirement of a precise natural unification, which is the inspiring principle of GUTs, 
we are inclined to prefer Higgs mechanism on the PS brane with a VEV near the cutoff scale,
as it gives the desired prediction for the sign of the Kaluza Klein contributions, restoring the correct value for 
the $\alpha_3(m_z)$ for a ration $\Lambda/M_c \gtrsim 100$. 

We have then performed a detailed analysis of the gauge coupling unification for all possible models, in which the
SO(10) gauge symmetry can be reduced to the Standard Model preserving the automatic D-T splitting,
including all next-to-leading effects. We have provided also a numerical confirmation of our result.
Our conclusion is that, as far as concerning the gauge coupling unification, Higgs mechanism on the PS brane 
is to be preferred to the one on the SO(10) preserving brane.      

\vspace*{1.0cm}
{\bf Acknowledgments}
We thank Ferruccio Feruglio for useful discussions and for his encouragement in our work. This project is partially
supported by the European Program MRTN-CT-2004-503369.

\end{document}